\documentclass[11pt,twoside]{article}
\usepackage{asp2004}
\usepackage{psfig}
\usepackage{epsf}
\usepackage{graphics}
\usepackage{lscape}
\def\edcomment#1{\iffalse\marginpar{\raggedright\sl#1\/}\else\relax\fi}
\reversemarginpar

\begin{document}
\title{Interplay between diffusion, accretion and  nuclear reactions in the atmospheres of
       Sirius and Przybylski's star}

\begin{quote}
  Alexander Yushchenko,$^{1,2}$ Vera Gopka,$^2$  Stephane Goriely,$^3$ Angelina Shavrina,$^4$
    Young  Woon Kang,$^1$
  Sergey Rostopchin,$^{5,6}$  Gennady Valyavin,$^{7,8}$
  David Mkrtichian,$^{1,2}$   Artie Hatzes,$^9$   Byeong-Cheol Lee,$^7$
  and Chulhee Kim$^{10}$\\
{\itshape $^1$Astrophysical Research center for the Structure and
              Evolution of the Cosmos, Sejong University, Seoul, 143-747, Korea}\\
{\itshape $^2$Astronomical observatory, Odessa National University, Odessa, 65014, Ukraine}\\
{\itshape $^3$Institut   d'Astronomie  et  d'Astrophysique,   Universit\'e  Libre  
              de Bruxelles, CP 226, 1050 Brussels, Belgium}\\
{\itshape $^4$Main Astronomical observatory, National Academy
              of Sciences of Ukraine, Kiev, 03680, Ukraine}\\
{\itshape $^5$The W.J. McDonald Observatory, University of Texas, Austin, TX 78712, USA}\\
{\itshape $^6$Crimean Astrophysical Observatoty, Nauchny, Crimea, 98409, Ukraine}\\
{\itshape $^7$Korea Astronomy and Space Science Institute, 61-1, Whaam-Dong, Youseong-Gu, Daejeon 305-348, Korea}\\
{\itshape $^8$Special Astrophysical Observatory, Russian Academy of Sciences,
              Nizhnii Arkhyz, Karachai Cherkess Republic, 369167, Russia}\\
{\itshape $^9$Thuringer Landessternwarte, Tautenburg, D-07778, Germany}\\
{\itshape $^{10}$Department of Earth Science Education, Chonbuk National University,
              Chonju, 561-756, Korea}\\
\end{quote}

\begin{abstract}
   The abundance anomalies in chemically peculiar B-F stars are usually
   explained by diffusion of chemical elements in the stable atmospheres
   of these stars.
   But it is well known that Cp stars with similar temperatures and gravities
   show very different chemical compositions.
   We show that the abundance patterns of several  stars can be
   influenced by accretion and~(or) nuclear reactions in stellar atmospheres.
   We report the result of determination of  abundances of elements in the atmosphere
   of hot Am star: Sirius~A  and show that Sirius~A was contaminated by s-process
   enriched matter  from Sirius~B (now a white dwarf).
   The second case is  Przybylski's star. The abundance pattern of
   this star is the second most studied one after the Sun with the
   abundances determined for about 60 chemical elements. Spectral lines
   of radioactive elements with short decay times were found in  the spectrum
   of this star. We report the results of investigation on the stratification
   of chemical elements in the atmosphere of Przybylski's star and the new identification of
   lines corresponding to short lived actinides in its spectrum.
   Possible explanations of  the abundances pattern
   of Przybylski's star (as well as HR465 and HD965) can be the natural radioactive decay
   of thorium, and uranium, the explosion of  a companion as a Supernova or
   nucleosynthesis events at stellar surface.
   \end{abstract}

\vspace{-0.5cm}
\section{Introduction}

   The diffusion of chemical elements in the atmospheres of late B -- early F main sequence
   stars is usually accepted as the major reason for the peculiar abundances
   in the atmospheres of   these objects. One of the biggest
   problems is the different abundance patterns for stars with very similar
   temperatures and gravities. It is clear that additional phenomena can influence
   the chemical composition. In the 70s, nuclear reactions
   in stellar atmospheres were discussed as the possible mechanism
   responsible for such chemical peculiarities.
   Proffitt \& Michaud (1989) pointed the possibility for these stars
   to have accreted $s$- or $r$-processes
   enriched matter.  They showed that
   one out of 500 peculiar stars can be expected to have
   surface abundance anomalies due to accretion from a binary companion
   that exploded as a Supernova (hereafter SN). Radiative diffusion is expected to  make these
   peculiarities  undetectable after a few millions years. No observational
   detection of this phenomena were found for B-F main sequence stars, but it
   was investigated in barium stars and permit to explain
   the overabundances of $s$-process elements.

   In this  paper we show the preliminary results of investigations of
   several Cp stars, namely Sirius~A,
   Przybylski's star, HD965, HR465. Accretion,
   natural radioactive decay of thorium and uranium,
   and nucleosynthesis in stellar atmospheres are discussed as the possible
   explanation of chemical composition of these objects.

\section{Observations and Data Analysis}

  The chemical abundances in the atmosphere of the main component of Sirius binary system
  are found using Rogerson (1987) atlas
  (spectral resolution 0.1 A, signal to noise ratio S/N$>$100, wavelength region 1649-3130 \AA)
  and spectrum obtained  at  2.6 meter telescope of McDonald observatory
  (spectral resolving power R=60,000, S/N$>$600 in red region, 3525-10200 \AA).
  The spectrum of Przybylski's  star (R=80,000, S/N$>$300 in red region, 3040-10400 \AA,
  UVES spectrograph)
  was taken  from the VLT archive (Bagnulo et al. 2002).
  Additional observations
  were made at 3.6 meter ESO telescope during 4 nights in March, 2004
  (R=110,000, 3780-6710 \AA, HARPS). Coaddition of
  13 spectra observed without iodine cell permit to reach S/N near 400-500 in red region.
  Observations of HD965 and HR465 were made at the 1.8 meter telescope
  of Boyhynsan observatory in Korea (R=80,000, S/N$>$150, 3780-9500 \AA, BOES).
  The abundances of the chemical elements were obtained using the spectrum synthesis method.
  The description of the codes and line lists can be found in Yushchenko et  al. (2004).


\section{Sirius}

 Sirius system consists of a main sequence star and a white dwarf (WD). The orbital period is
 about 50 years. The atmosphere of the main sequence star can be
 contaminated by s-process enriched matter at the time the WD was a red giant.
 The contamination of Sirius~A by the companion was pointed by Lambert et al. (1982) as
 a reason of anomalous CNO abundances.
 We calculated the abundances of elements with atomic numbers Z$\ge$29 in the
 atmosphere of Sirius~A. Atlas12 atmosphere
 model was built with parameters T$_{eff}$=9900, logg=4.3,
 microturbulent velocity 1.85 km/s, and Qiu et al. (2001) abundances.
 Rotational velocity was accepted to be equal $vsini$=16~km/s.
 We estimated the abundances of Cu~I (1 spectral line), Zn~I (3), Sr~II (2), Y~II (11),
 Zr~II (13), Ba~II (5), La~II (2), Ce~II (11), Pr~III (1), Nd~III (2), Gd~II (1), Os~II (1),
 Pb~II (1) using McDonald observatory spectrum and abundances of
 Ga~II (1 spectral line),  Mo~II (1), Cd~II (1), Hf~II (1), W~II (1), Re~II (2),
 Os~II (4), Pt~II (2), Hg~II (1), Pb~II (2) using Copernicus UV spectrum.
 Preliminary results are shown in Fig. 1 where
 the contamination of the atmosphere is clearly seen. As it was noted by Proffitt and
 Michaud (1989) the peculiarities of this type should be destroyed by diffusion in few
 millions years.

  We can point three estimates of the age of Sirius WD:
 near 160$\cdot$10$^6$ years (standard theory of WD: Holberg et al. 1998),
 less than 1-5$\cdot$10$^6$  years  (our abundance analysis and diffusion theory),
 and  1-2$\cdot$10$^3$~ years
  (historical researches: See 1927, Whittet 1999 and references therein).

 \setcounter{figure}{0}
 \begin{figure}[!h]
 \plotfiddle{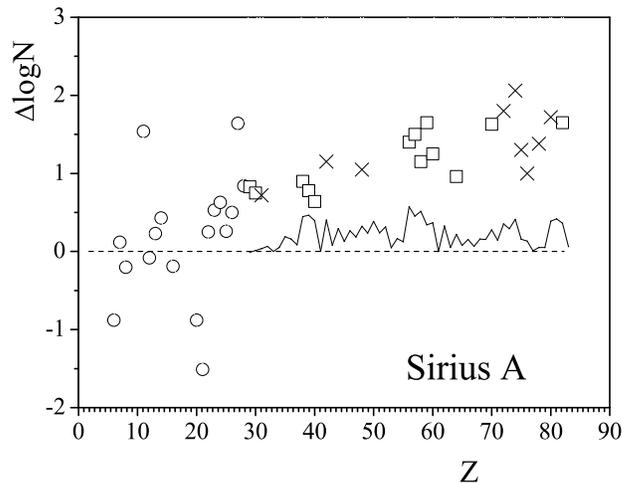}{55mm}{}{90}{90}{-140}{-25}
 \caption{Chemical composition of
  Sirius~A. The axes are atomic numbers and logarithmic abundances
  with respect to solar values.  Abundances of elements with Z$<$29 (from
  Qiu et. al (2001)) are marked by  circles.
  Our results are shown by squares (values obtained from visual spectrum)
  and crosses (from UV spectrum).
  The line corresponds to calculated s-process-rich abundances obtained from the
  wind accretion model for mild barium star $\zeta$ Cyg A (Yushchenko et al. 2004).
  The extremums of this curve coincide with extremums of
  observed  abundances}
 \end{figure}

 To check the third possibility,  we tried to find the lines of Tc (Z=43) in the
 optical and UV spectra of Sirius~A. No sign of Tc lines was found.
 So, if Tc was produced at the final stages of evolution of red giant,
 the age of WD should significantly exceed the decay time of the Tc isotopes,
  i.e about 10$^5$ years.

  The discrimination between the first two cases needs additional investigation.
  If the contamination should be destroyed by diffusion, what is the
  final abundance pattern?
  The contaminated abundances are observed in barium stars. Is it possible to observe
  similar patterns in Sirius type binaries with age of WD more than few millions  years?

\section{Przybylski's star (HD101065)}

   Cowley et al. (2004)
   found the lines of Tc  and Pm  in the spectra of Przybylski's star (hereafter PS)  and HD965.
   Gopka et al. (2004)
   and Bidelman (2005) found the lines of short-lived elements with atomic numbers
   83$<$Z$<$100 in the spectrum of PS.
   The lines of all elements except At and Fr
   (Z=85 \& 87) were identified in both studies. Gopka et al. (2004) proposed
   some explanations for the existence of these elements in the stellar atmospheres.
   Gopka et al. (2006) reported the identification  of  Tc, Pm,  and lines of short
   lived elements with  83$<$Z$<$100 in the spectra of PS, HR465, and HD965.
   The above-mentioned investigations were based on the NIST  wavelengths data for
   elements with Z$>$83 (Sansonetti et al. 2004).
   Of course,  these identifications needs additional justifications.

   The number of unidentified lines in the spectra of
   PS and similar stars is of the order of several lines per angstrom.
   In the 70s of the last century there was a hope that new laboratory data for
   lanthanides will help to identify these lines. After 3 decades
   we have a lot of additional line data. The largest new set of lanthanides lines
   is  DREAM database (Biemont et al. 2002a). But available lines of stable
   elements are still not sufficient for identification of unknown lines.
   If unstable elements exist in the atmosphere of PS it is quit natural to expect that
   the analysis of line lists of these elements will permit to find new identifications.

    We tried to find new lines using  atomic spectra of actinides by
   Blaise \&  Wyart (2005).  52  new  lines of
   unstable actinides are found in the spectrum of PS.
   Oscillator strengths for these lines are not known, even ionization stages
   are not pointed  for einsteinium.
   Let us consider only lines of the second spectra of the elements
   with at least four identifications in the spectrum of PS. These are
   10, 6, 23, 5, and 4  lines of Ac~II, Pa~II, Pu~II, Am~II, and Bk~II respectively.
   Note, that 23 lines is found for Pu. The decay time of ~$^{244}$Pu is
    80$\cdot$10$^6$ years -- the longest  among actinides,
   after Th and U, and not negligible in comparison with
   stellar evolution time scale. One of Pu lines is shown in Fig. 3.

  Three  scenarios can be called for to explain the existence of short-lived
  radioactive elements  in stellar atmosphere.
  Gopka et al. (2004) pointed that due to diffusion in stellar
  atmospheres layers with Th and U overabundances should exist. Due to the Th and U natural
  radioactive decay  elements with 83$<$Z$<$92 can be produced. Elements with
  Z$>$92 are created  by reactions on lighter  elements  with
  neutrons and $\alpha$-particles produced in the decay chains.
  All these elements are found in terrestrial radioactive ores.
  The ores with concentration of Th
  and U near one percent are considered as the best ores for industry.

 \begin{figure}[!h]
 \plotfiddle{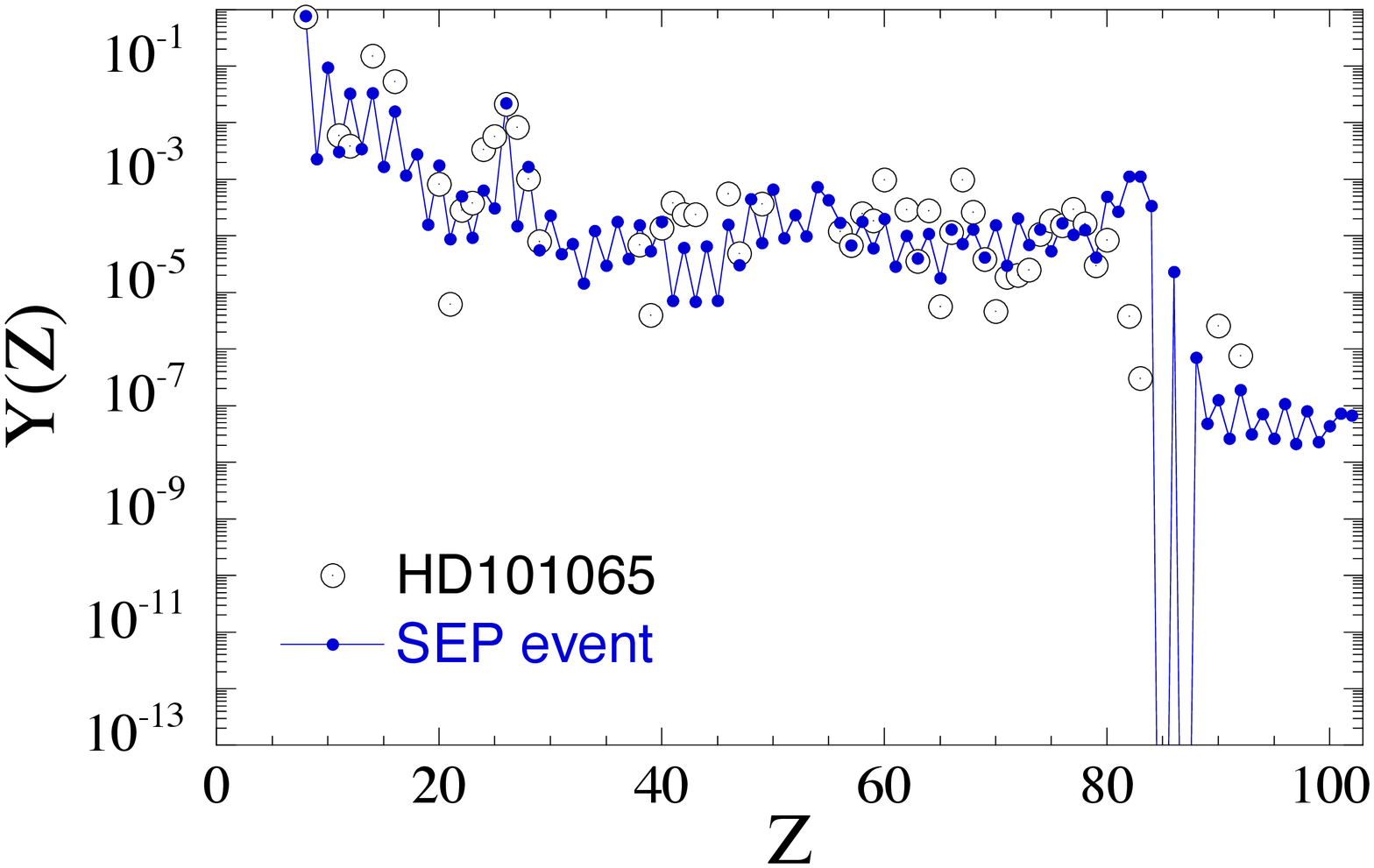}{45mm}{}{40}{35}{-180}{-35}
 \caption{Comparison of Cowley et al. (2000) abundances
 (circles) with those obtained for one of the possible nucleosynthesis events
  (see text).}
 \end{figure}

  Abundance pattern of PS can be fitted by scaled  solar $r$-process pattern (Cowley et al. 2000).
  It goes along the old idea  that the PS abundances
  originate from the SN explosion of its binary component.
  Gopka et al. (2004) found the upper limits of Pb and Bi abundances.
  These limits are smaller than  scaled  $r$-process values.
  Kuchowicz (1973) pointed, that it is a sign of recent SN event.

 \begin{figure}[!h]
 \plotfiddle{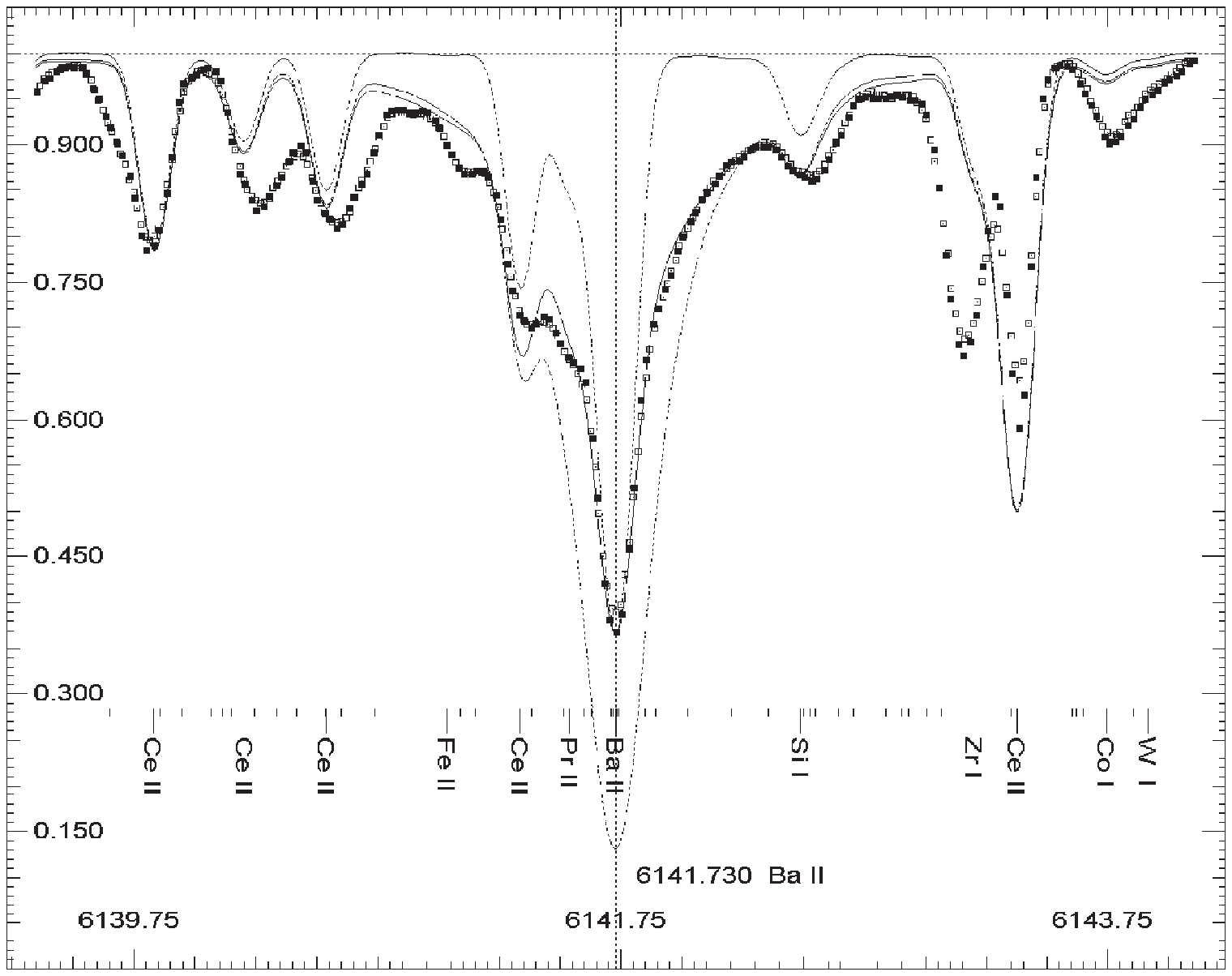}{51mm}{}{ 88}{50}{-267}{-200}
 \plotfiddle{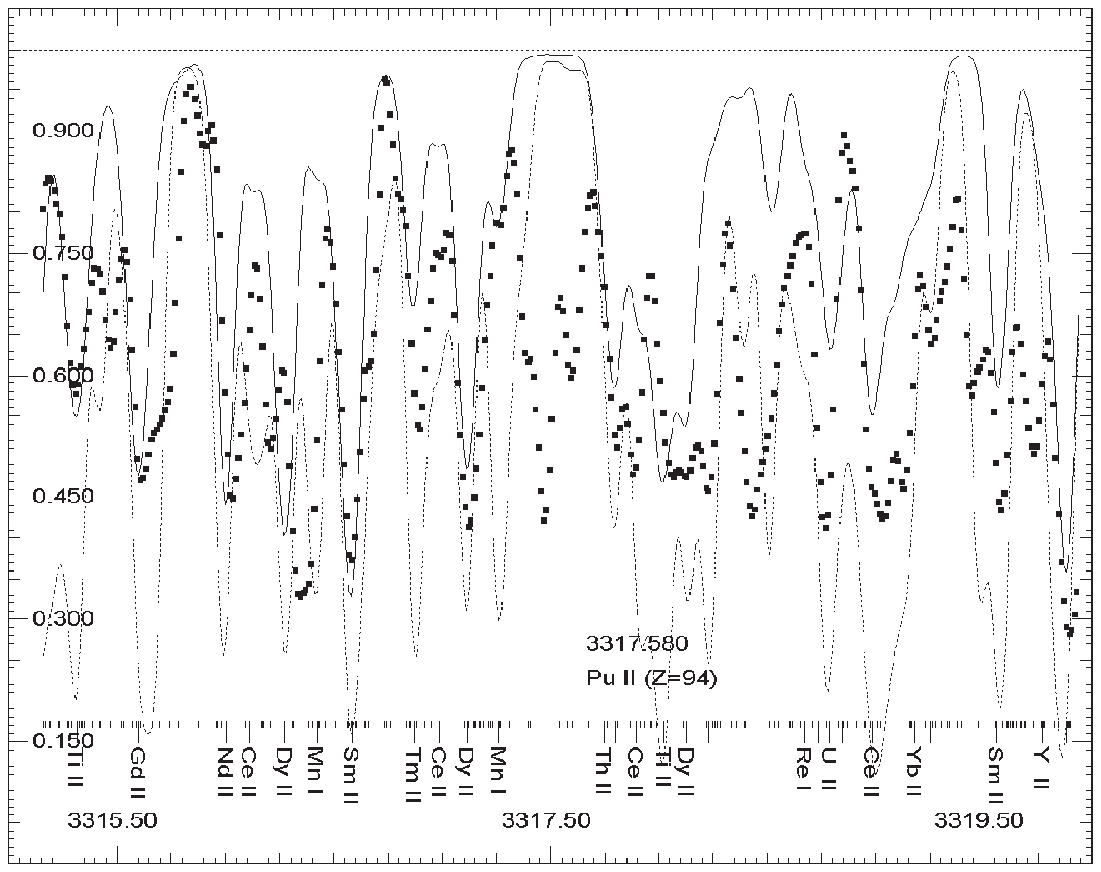}         {60mm}{}{122}{71}{-368}{-345}
 \caption{
  {\itshape Upper:\/}  Spectrum of PS in the vicinity of Ba~II line 6141~\AA.
   Filled and open squares -- VLT and HARPS spectra. For this and other spectral regions
   the difference between the observed  spectra is within the noise level.
   Three synthetic spectra
   are shown. Two spectra are calculated with flat barium distribution using
  $N(Ba)/N_{total}$=10$^{-7.65}$ and 10$^{-9.95}$ to fit the wings and
  the core.
   The third one --  with stratified barium abundances to fit the profile  of the line.
  {\itshape Bottom:\/}
  Spectrum of PS in the vicinity of Pu~II line  $\lambda$3317.58~\AA.
  Points - VLT spectrum. Lines --
  synthetic spectra. Dotted line -- Cowley at al. (2000) flat abundances,
  solid line -- stratified abundances (preliminary values).
  Shavrina et al. (2003) atmosphere model is used.
  Part of the strongest lines  are marked.}
 \end{figure}

   Finally, nuclear reactions at the stellar surface can naturally be called
   for to explain the formation of
   radioactive elements. The large magnetic field observed in
   Ap stars can be the origin of a significant acceleration of
   charged  stellar energetic particles (SEP), mainly protons and  $\alpha$-particles,
   that in turn can by
   interaction with the stellar material modify the surface content. External
   sources from cosmic rays or jet-like accelerated particles could also be considered.
   Due to the
   unknown characteristics of the accelerated particles, a purely parametric
   approach was followed, taken as free parameters the proton and $\alpha$-particle
   flux amplitude and energy distribution, and the time of irradiation.  To
   estimate the resulting nucleosynthesis, a nuclear reaction network including all
   nuclei heavier than oxygen  up to $Z=102$ and located between the proton drip
   line and the neutron-rich side of the valley of stability is used. Proton,
   $\alpha$  and neutron captures, as well as $\alpha$-, $\beta$- and spontaneous
   fission decays are considered. This includes some 240000 reactions on 3940
   different species.  The initial abundance distribution is assumed to be
   solar. Our calculation shows that many aspects of the composition of PS
   and other chemically peculiar stars can be explained.
   As an example, we compare in Fig.~2 the PS abundances with those obtained for
   a given irradiation event characterized by
   a proton flux of $\Phi_p=100$~mb$^{-1}$\,s$^{-1}$, and
    $\alpha$-particle flux $\Phi_{\alpha}/\Phi_p=0.2$,
     both with energies per nucleon ranging uniformly
     between 40 and 50~MeV. The irradiation time is 24 min.
      Note that we also consider here the possibility that
       the observed abundances is a mix (for example by diffusion)
        between  the irradiated material and part of the ambient
         unaffected surface matter. For a 3-to-1 ratio between
          unaffected and irradiated material, the pattern show
           in Fig.~2 is obtained. As can be seen,   a significant
           production of $Z>30$ heavy elements is found, and can
            be explained through  secondary neutron captures.
             Another attractive feature of the proposed
   process is the systematic production of Tc and Pm, and the possible production
   of actinides and subactinides, as suggested by our observations. More details
   on these calculations will be published somewhere else.

 \begin{figure}[!h]
 \plotfiddle{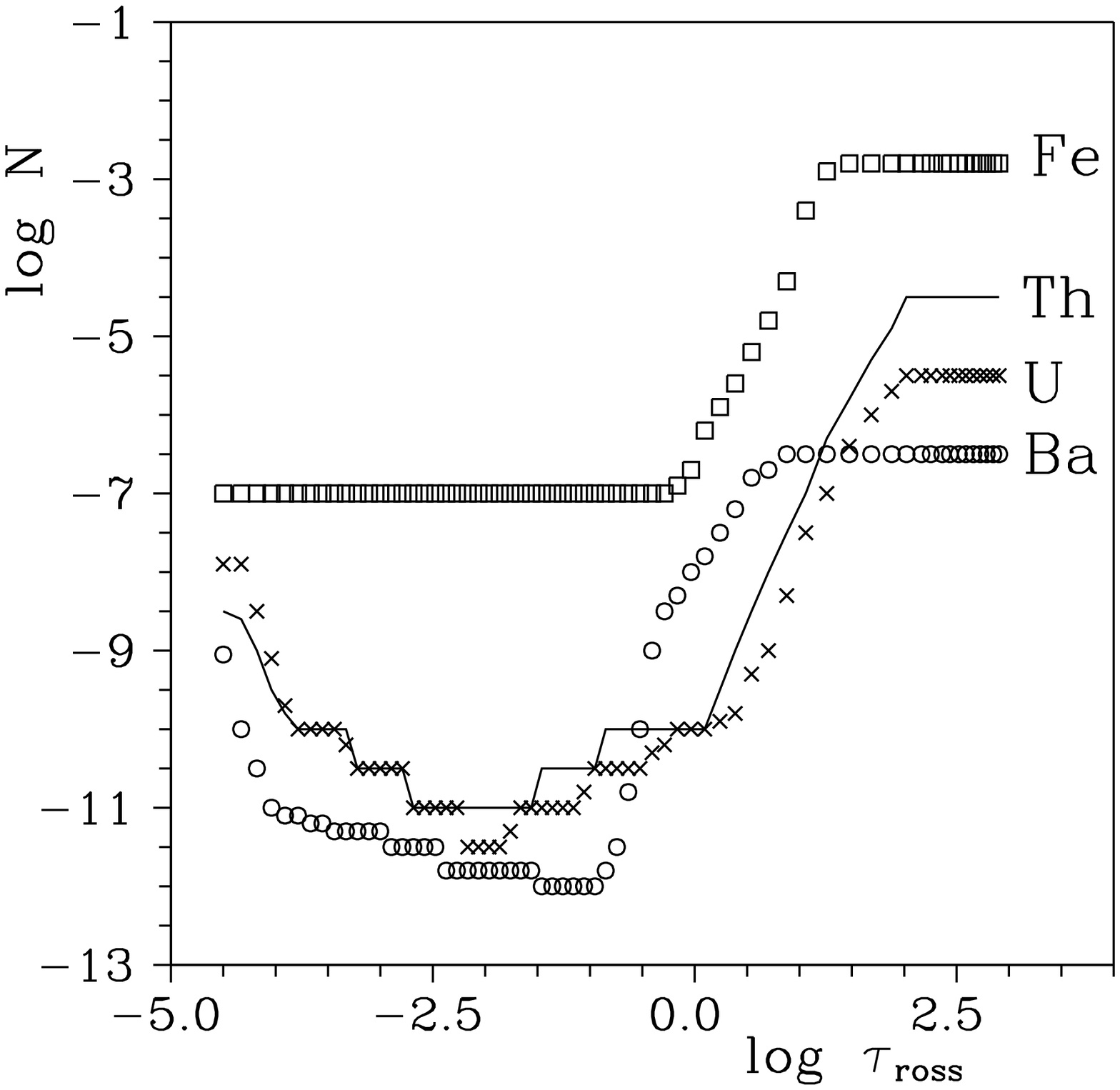 } { 49mm}{}{36}{36}{-200}{-35}
 \plotfiddle{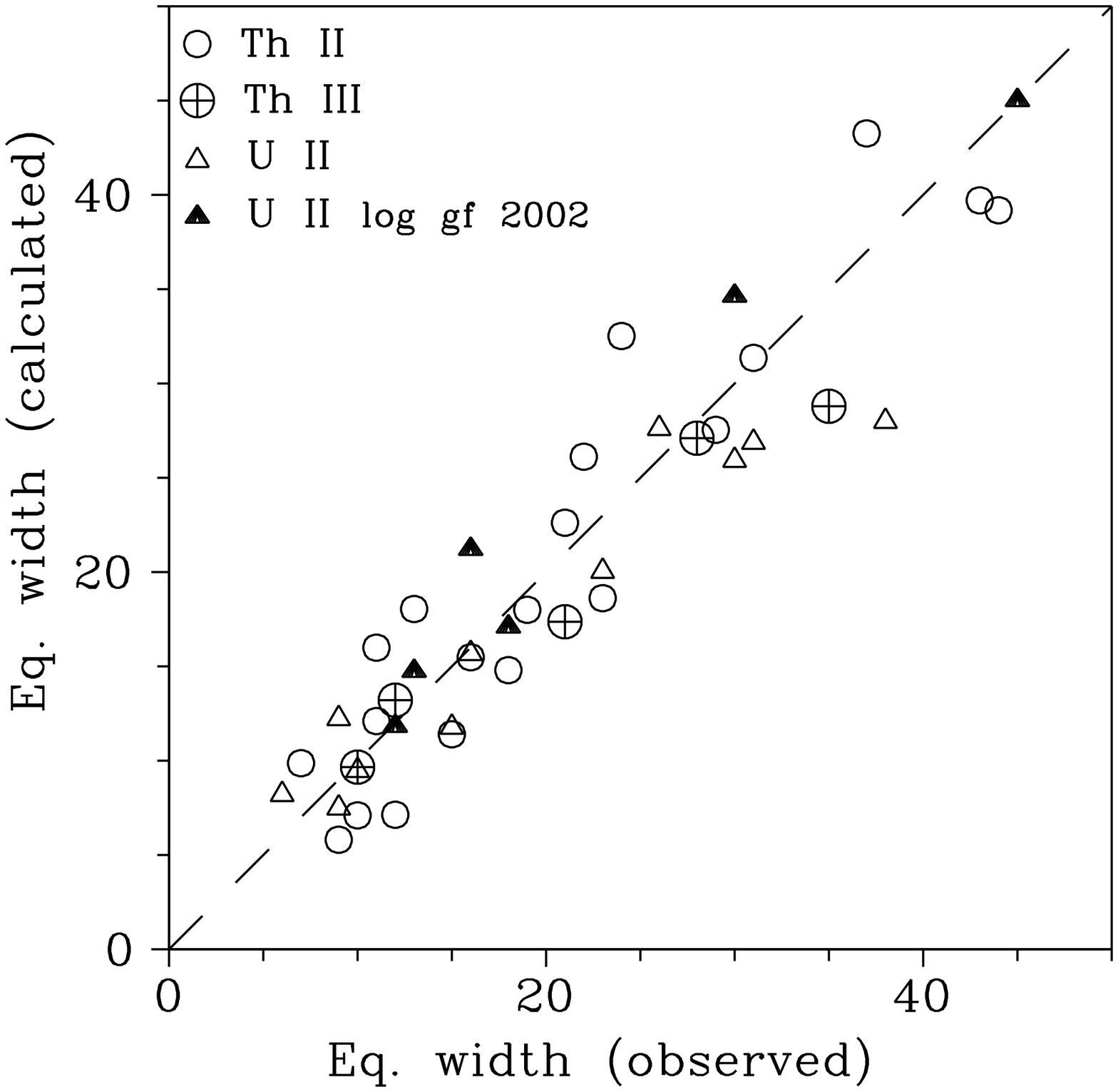} {-49mm}{}{36}{36}{-008}{-10}
 \plotfiddle{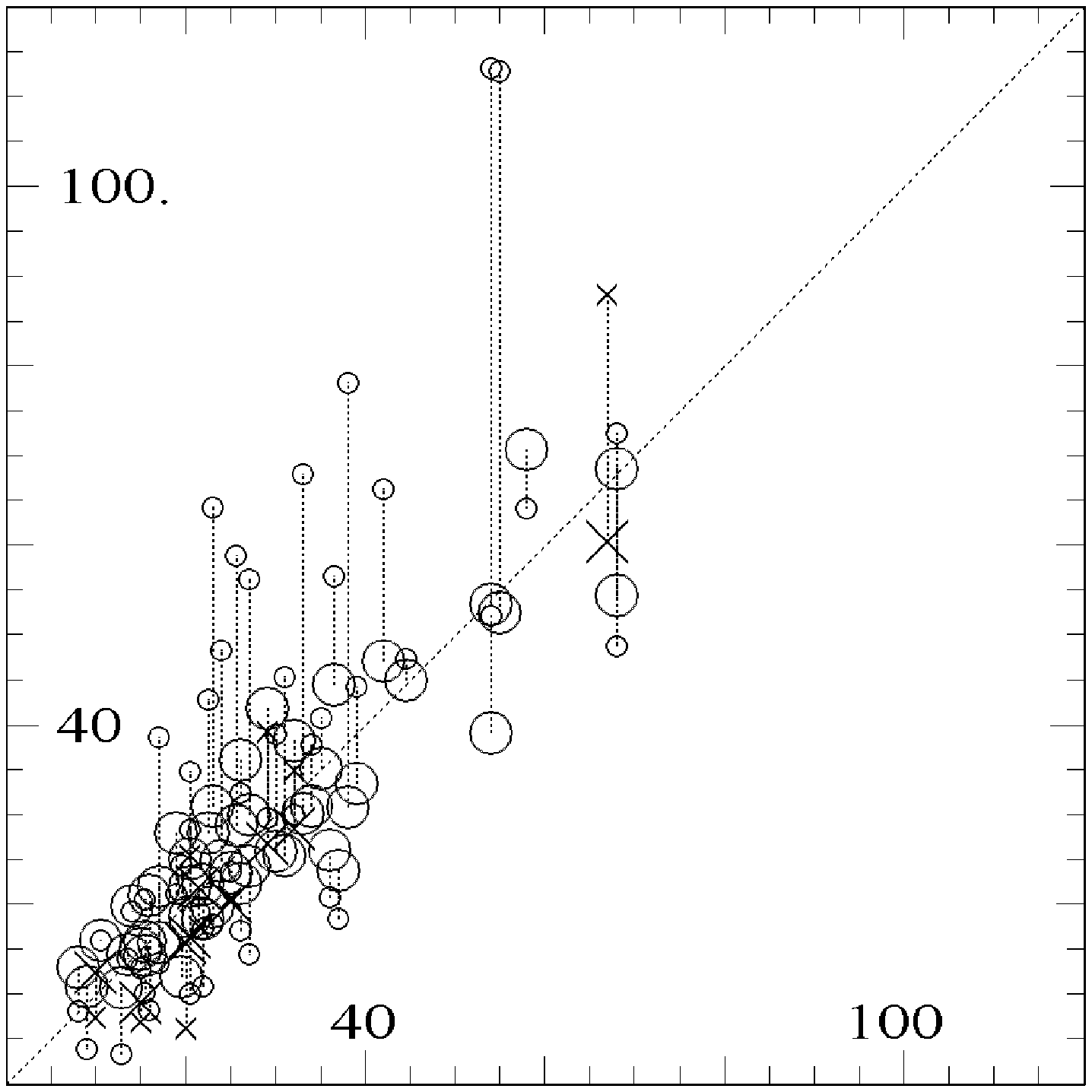} {-45mm}{}{38}{38}{-192}{-160}
 \plotfiddle{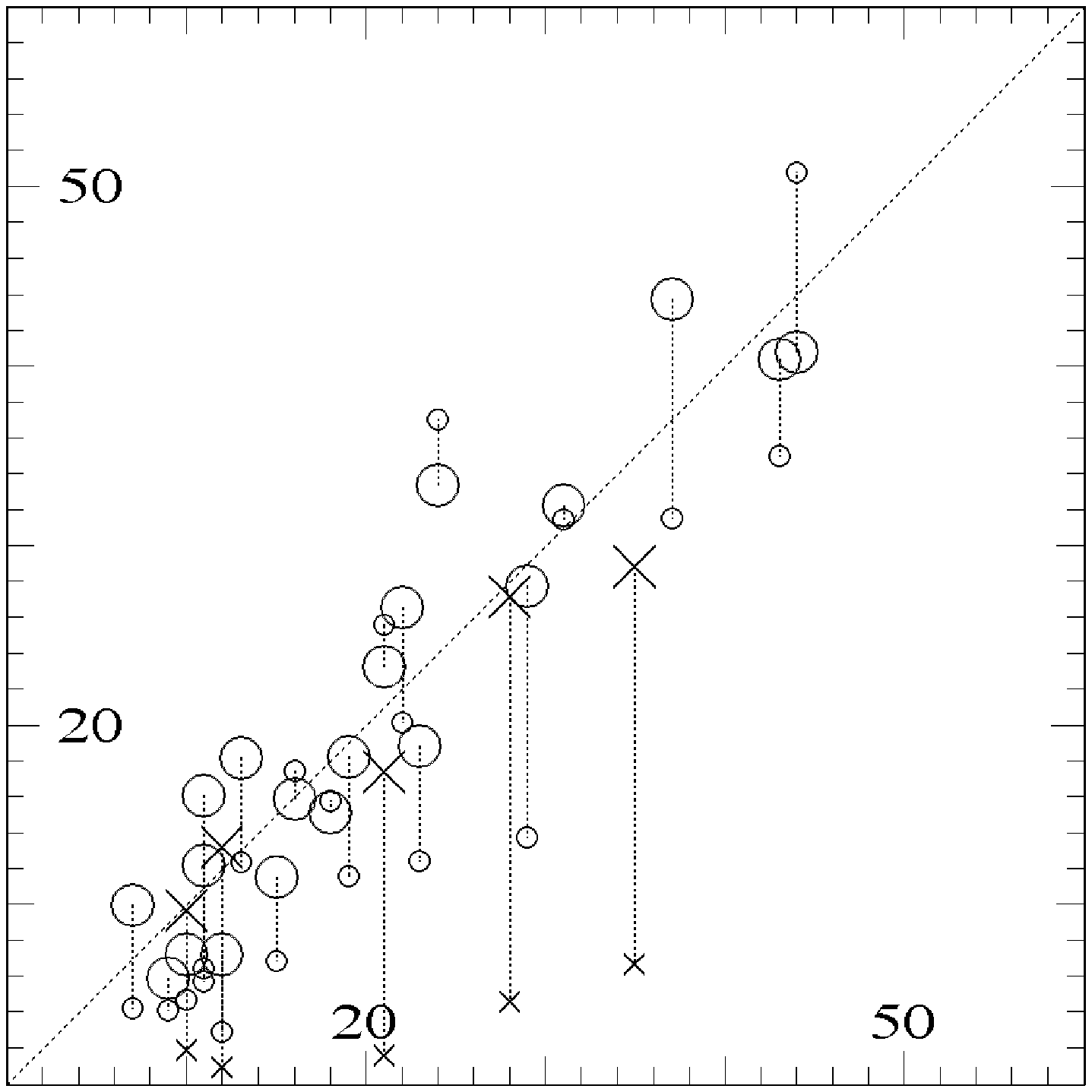}{ 45mm}{}{38}{38}{ -00}{-20}
 \caption{{\itshape Upper left:\/} Stratification of
  Fe, Ba, Th, and  U in the atmosphere of
  PS. The axes are optical depth and abundance in the scale logN(H)=0 .
 {\itshape Upper right:\/} Comparison of observed equivalent widths of Th and U lines
  in the spectrum of PS
  (in m\AA) with  theoretical values calculated with stratified abundances.
  Oscillator strengths are taken
  from Nilsson et al. (2002a) for Th~II,
  from Biemont et al. (2002b) for Th~III,
  from Nilsson et al. (2002b) for  U~II.
  We used 17 lines of U~II, for 6 lines
  oscillator strengths from Nilsson et al. (2002b) are available.
  38 out of the 45 Th and U lines have wavelengths longer than 5000~\AA, 7 lines are
  in the 4700-5000 \AA~ wavelength region.
  {\itshape Bottom left:\/} Comparison of observed and calculated equivalent widths
  of Fe lines. Calculated values  are shown for flat
  and stratified (small and large symbols) iron distribution. Dotted lines connect the points
  calculated for flat and stratified abundances.
  Circles - neutral iron, crosses -   ionized iron.
  {\itshape Bottom right:\/} The same for Th.
  Circles and crosses -- second  and third spectra.}
 \end{figure}

  To select one of these scenarios we need the data on stratification and
  abundances of chemical elements in the atmosphere of PS.
  We found the stratification of barium using the profile of the strong Ba~II line
  at $\lambda$~6141~\AA~ (Fig. 3). Similar result for barium was published
  by Ryabchikova et al. (2003). It should be noted that Cowley et al. (2000)
  abundances and atmosphere model with flat abundances can fit the observed
  spectrum in red and visual wavelength regions but fail in near UV (Fig. 3).
  The stratification of chemical elements in the atmosphere of PS
  is very plausible hypothesis to reach an agreement.

  Analysis of equivalent widths of Fe, Th, and U permits to find the distribution of
  these elements in the atmosphere of PS (see Fig. 4).
  Test calculations show  an evidence of strong  stratification for
  iron group elements,
  lanthanides and other elements in the atmosphere of PS.
  The influence of such a stratification on the spectrum is shown in Fig. 3.
  The abundances of practically all elements are increased at the bottom of the atmosphere.
  Abundances can differ by several orders of magnitudes at different optical
  depths, reaching up to 6 dex for Th in particular.
  The abundances of Th at the bottom of atmosphere of PS is near $N/N_{total}$=10$^{-4.5}$,
  i.e  only two orders of magnitude lower, than that in the best radioactive ores.
  So, spectroscopic registration of short-lived isotopes in the spectrum of PS seems
  reliable, the abundances of these isotopes should be several orders lower than the
  abundances of Th \& U.
  The oscillator strengths for actinides lines (except Th \& U) are not known, but
  it is possible to estimate low limits of abundances.
  If we set the logarithms of oscillator strengths of identified
  lines of second spectra of
  Ac, Pa, Pu, Am, and Bk  to be zero,
  the mean $log(gf\cdot\epsilon)$ values are equal 1.2, 1.7, 1.6, 1.3, 1.1 respectively.
  Similar values for Th and U are 1.1 and 1.1 (scale logN(H)=12 is used). So
  the abundances of unstable actinides  can be close to that of Th \& U.
  It is the strong argument to choice second or third of above mentioned hypotheses.
  The first one can be true only if the heaviest stable elements in stellar atmospheres
  are not Th \& U, but that  near island of stability,
  as it  was proposed by Kuchowicz (1973).

 All the three scenarios described here still need to be investigated in more details.
 The combination of different physical mechanisms, like nuclear reaction and
 diffusion, may also have contributed to the presently observed stellar surface of HD101065
 and similar stars.


\vfill\pagebreak


\begin{thebibliography}{}
%
\bibitem{ay}  Bagnulo, S., Jehin, E., Ledoux, C.,  Cabanac,  R.,  Melo, C.,  Gilmozzi,  R.
    and the ESO Paranal Science Operations Team 2002,
    A Library of High-Resolution Spectra of Stars across the Hertzsprung-Russell Diagram ,
    $http://www.sc.eso.org/santiago/uvespop/index.html$
%
 \bibitem{1}  Bidelman, W.~P. 2005,  in
               Cosmic Abundances as Records of Stellar Evolution and Nucleosynthesis
               in honor of David L. Lambert,
               ed. T.~G. Barnes III, \& F.~N. Bash,
               Astron. Soc. of the Pacific Conf. Ser. 336,
               San Francisco,  309
%
\bibitem{a1}   Biemont, J., Palmeri, P., Quinet, P. 2002a,
               Database on Rare Earths At Mons University,
               http://www.umh.ac.be/~astro/dream.shtml
%
\bibitem{ay}  Biemont, E.,  Palmeri, P., Quinet, P.,  Zhang, Z.~G., \&  Svanberg, S.
              2002b, \apj, 567, 1276
%
 \bibitem{2a} Blaise, J., \&  Wyart, J.-F. 2005, Energy Levels and Atomic Spectra of
               Actinides, $http://www.lac.u-psud.fr/Database/Contents.html$
%
 \bibitem{2}  Cowley, C.~R., Ryabchikova, T., Kupka, F., Bord, D.~J.,
              Mathys, G., \&  Bidelman, W.~P. 2000,  \mnras, 317, 299
%
 \bibitem{2a} Cowley, C.~R., Bidelman, W.~P., Hubrig, S., Mathys, G., \&  Bord, D.~J.
              2004, \aap, 419, 1087
%
 \bibitem{3}  Gopka, V.~F.,  Yushchenko, A.~V.,  Shavrina, A.~V.,
              Mkrtichian, D.~E.,  Hatzes, A.~P.,  Andrievsky, S.~M., \&  Chernysheva, L.~V.
              2004,
              in The A-Star Puzzle, IAU Symp. 224,  ed. J. Zverko, J. Ziznovsky,
              S.~J. Adelman, and W.~W. Weiss, 734
%
 \bibitem{3a} Gopka, V.,  Yushchenko, A., Goriely, S.,  Shavrina,  A., \&  Kang Y.~W.
              2006,
              in Origin of Matter and Evolution of Galaxies 2005, ed. S. Kubono
              (accepted for publ.)
%
%
%
 \bibitem{4a}  Holberg, J.~B.,  Barstow, M.~A.,  Bruhweiler, F.~C.,
                Cruise, A.~M., \&  Penny, A.~J. 1998,   \apj, 497, 935
%
 \bibitem{7}  Kuchowicz, B. 1973, Quartely J. Royal Aston. Soc., 14, 121
%
 \bibitem{7a} Lambert, D.~L., Roby, S.~W., \& Bell R.A.  1982, \apj, 254, 663
%
%
\bibitem{ax}  Nilsson, H.,  Zhang, Z.~G., Lundberg, H., Johansson, S., \&   Nordstrom, B.
              2002a, \aap, 382, 368
%
\bibitem{az}  Nilsson, H.,  Ivarsson, S., Johansson,  S., \&  Lundberg, H.,
              2002b, \aap, 381, 1090
%
 \bibitem{6}  Proffit, C.~R., \&  Michaud, G. 1989,  \apj, 345, 998
%
 \bibitem{6q} Qiu, H.~M., Zhao, G., Chen, Y.~Q.,  \&  Li, Z.~W. 2001,  \apj, 548, 953
%
 \bibitem{7}  Rogerson, J.~B. 1987, \apj, 163, 369
%
 \bibitem{7a} Ryabchikova, T.~A.,  Wade, G.~A., \&  LeBlanc, F.   2003,
              in Modelling of Stellar Atmospheres, IAU Symp. 210,
              ed. N. Piskunov, W.~W. Weiss, \& D.~F. Gray,  301.
%
 \bibitem{7b}  Sansonetti, J.~E.,  Martin, W.~C., \&  Young, S.~L. 2004,
               Handbook of Basic Atomic Spectroscopic Data.
               http://physics.nist.gov/PhysRefData/Handbook/index.html
%
%
%
 \bibitem{8}  See, T.~J. 1927,  Astr. Nachr., 229, 245
%
 \bibitem{15}  Shavrina, A.~V.,  Polosukhina, N.~S.,  Pavlenko, Ya.~V.,
              Yushchenko, A.~V.,  Quinet, P.,  Hack, M.,  North, P.,
              Gopka, V.~F.,  Zverko, J.,  Zhiznovsky, J., \&  Veles, A. 2003,
             \aap, 409, 707
%
%
 \bibitem{9}  Whittet, D.~C. 1999,  \mnras, 310, 355
%
%
  \bibitem{12}  Yushchenko, A.~V.,
               Gopka, V.~F.,  Kim, C.,  Liang, Y.~C., Musaev, F.~A., \&  Galazutdinov, G.~A.,
               2004,  \aap, 413,   1105
%
%
%
%


\end{thebibliography}
\end{document}